\documentclass[fleqn,twoside]{article}
\usepackage{gc,epsf}
\usepackage [russian]{babel}
\heads{Olga V. Babourova}
      {Modified Friedmann--Leme\^{\i}tre Equation for Dilaton-Spin Dark Matter
       in Weyl--Cartan Space}

\begin{document}
\twocolumn[
\Arthead{10}{2004}{1-2 (37-38)}{121}{126}

\Title {MODIFIED FRIEDMANN--LEMA\^ITRE EQUATION \yy
FOR DILATON-SPIN DARK MATTER IN WEYL--CARTAN SPACE}

 \Author{Olga V. Babourova\foom 1}
 {Moscow State University, Faculty of Physics, Department of Theoretical Physics,\\
  Leninskie Gory, Moscow 119992, Russian Federation}

\Abstract
    {It is proposed to consider dark matter as a perfect dilaton-spin fluid
(with particles endowed with intrinsic spin and dilaton charge) in
the framework of a gravitational theory with a Weyl--Cartan geometrical structure.
The modified Friedmann--Lema\^{\i}tre  equation (with a cosmological term) is obtained
for the homogeneous and isotropic Universe filled with the dilaton-spin dark matter.
On the basis of this equation, we develop a nonsingular cosmological model starting from
an inflation-like stage (for super-stiff equation of state), passing radiation-dominated
and matter-dominated decelerating stages and turning into a post-Friedmann accelerating era.}

]
\email 1 {baburova@orc.ru}

\section{Introduction}
\setcounter{equation}{0}
The basis concept of modern fundamental physics consists in the
preposition that the spacetime geometrical structure is compatible with the
pro\-per\-ties of matter filling the spacetime. It means that matter dynamics
determines the metric and connection of the spacetime manifold and
is in turn  determined by the spacetime geometric pro\-per\-ties. Therefore the
possible deviation from the geometrical structure of general relativity
should be stipulated by the existence of matter with unusual
properties, which fills spacetime, generates its structure and interacts
with it. As examples of such matter, perfect media with in\-trin\-sic degrees
of freedom are considered, such as a perfect fluid with spin and
non-Abelian colour charge \cite{bfdan}, a perfect hypermomentum fluid (see
\cite{Arg}--\cite{bfhyp} and references therein), and a perfect dila\-ton-spin
fluid \cite{Mod}. All these fluids are generalization of the
Weyssenhoff--Raabe perfect spin fluid \cite{WR}.
\par
Modern observations \cite{nabl,Perl} lead to the conclusion
on the existence of dark (nonluminous) matter whose density
exceeds by an order of magnitu\-de the density of baryonic matter which
forms stars and the luminous components of galaxies. It is dark
matter interacting with positive vacuum energy or quintessence (whose density has
the same order of magnitude)  \cite{Os-Perl,St} that realizes the dynamics
of the Universe in the modern era. Another important consequence of the modern
observations is an understanding of the fact that the end of the Friedmann
era occurs when decele\-ra\-ted expansion is followed by accelerated expan\-si\-on,
a transition to an unrestrained exponential expansion being possible.
\par
The hypothesis on the existence of dark matter in galaxies was proposed by
Zwicky in his pione\-er\-ing work \cite{Zwic}. But the essence of dark matter is
yet unknown. The hypothesis that dark matter was endowed with a new kind of
gravitational charge which generates a short-range gravitational
interaction of Proca type was put forward in \cite{Tuck}. This interaction is best
appreciated in terms of the existence of Weyl--Cartan spacetime. Independent\-ly,
in \cite{GC-99} it was shown that Weyl--Cartan geometry was generated by a
perfect dilaton-spin fluid, and the corresponding non-singular cosmological
model was constructed. Then, in \cite{CQG-03} the hypothe\-sis on a perfect
dilaton-spin fluid as a model of dark matter was proposed in a gravitational
theory with Weyl--Cartan geometrical structure.
\par
    Within the framework of these ideas, a modified Friedmann--Leme\^{\i}tre (FL)
equation with a cosmo\-lo\-gical term for a homogeneous and isotropic Universe
filled with the dilaton-spin dark matter was constructed for an arbitrary equation
of state of dark matter \cite{GC-99}--\cite {GC-03}. In the present paper,
solutions of this equation for various equations of state are obtained. An
inflation-like solution is obtained for superstiff equation of state
of dark matter at the early stage of the Universe. The evolution of the Universe
starts from a very small but non-zero size, then passes Friedmann decelerating stage
and turns into a post-Friedmann accelerating era.
\par
    Throughout the paper the signature of the metric is assumed to be
$(+,+,+,-)$, and the conventi\-ons $c = 1$, $\hbar = 1$ are used.

\section{Weyl--Cartan space}
\setcounter{equation}{0}
 Let us consider a connected 4-dimensional
oriented differentiable manifold ${\cal M}$ equipped with a metric
$\breve g$ of index 1, a linear connection $\Gamma$ and a volume
4-form $\eta$. Then a Weyl--Cartan space $CW_4$ is defined as the
space equipped with the curvature 2-form ${\cal R}^{\alpha}\!_{\beta}$
and the torsion 2-form ${\cal T}^{\alpha}$ with the metric tensor and
the connection 1-form obeying the condition
\begin{eqnarray}
&&- {\rm D} g_{\alpha\beta} = {\cal Q}_{\alpha\beta} = \frac{1}{4}
g_{\alpha\beta}{\cal Q}\; , \nonumber \\
&& {\cal Q}= g^{\alpha\beta}{\cal Q}
_{\alpha\beta} = Q_{\alpha}\theta^{\alpha}\; ,
\end{eqnarray}
where ${\cal Q}_{\alpha\beta}$ is the nonmetricity 1-form, ${\cal
Q}$ the Weyl 1-form, and ${\rm D} = {\rm d } +
\Gamma\wedge\ldots$ the exterior covariant differential. Here
$\theta^\alpha$ ($\alpha = 1,2,3,4$) is cobasis of 1-forms and $\wedge$
is the exterior product operator.
\par
The curvature  2-form ${\cal R}^{\alpha}\!_{\beta}$ and the torsion 2-form
${\cal T}^{\alpha}$,
\begin{eqnarray}
&&{\cal R}^{\alpha}\!_{\beta}=\frac{1}{2}R^{\alpha}\!_{\beta\gamma\lambda}
\theta^{\gamma}\wedge\theta^{\lambda}\,, \nonumber \\
&& {\cal T}^{\alpha}=\frac{1}{2} T^{\alpha}\!_{\beta\gamma}\theta^{\beta}
\wedge\theta^{\gamma}, \quad T^{\alpha}\!_{\beta\gamma} =
-2\Gamma^\alpha\!_{[\beta\gamma ]},\nonumber
\end{eqnarray}
are defined by the Cartan structure equations,
\begin{eqnarray}
&&{\cal R}^{\alpha}\!_{\beta}={\rm
d}\Gamma^{\alpha}\!_{\beta}+\Gamma^{\alpha}\!_
{\gamma}\wedge\Gamma^{\gamma}\!_{\beta}\,, \nonumber \\
&& {\cal T}^{\alpha} = {\rm d}\theta^{\alpha}+\Gamma^
{\alpha}\!_{\beta}\wedge\theta^{\beta}\,.\nonumber
\end{eqnarray}
The Bianchi identities for the curvature 2-form, the torsion
2-form and the Weyl 1-form are valid,
\[
{\rm D} {\cal R}^{\alpha}\!_{\beta} = 0\,, \;\;  {\rm D} {\cal
T}^{\alpha} = {\cal R}^{\alpha}\!_{\beta}\wedge \theta^\beta \,,
\;\; {\rm d}{\cal Q} = 2{\cal R}^{\gamma}\!_{\gamma} \,.
\]
\par
  The torsion 2-form can be decomposed into the sum of irreducible pieces
$\stackrel{(i)}{{\cal T}}\!^{a}$: the traceless 2-form with $i=1$, the
trace 2-form with $i=2$ and the pseudotrace 2-form with $i=3$
\cite{He:pr, Ob-Hel},
\[
{\cal T}^{\alpha} = \stackrel{(1)}{{\cal T}}\!^{\alpha} + \stackrel{(2)}
{{\cal T}}\!^{\alpha} + \stackrel{(3)}{{\cal T}}\!^{\alpha}\; ,
\]
where the torsion trace 2-form and the torsion pseudotrace 2-form are
determined by the expres\-sions,
\begin{eqnarray}
\stackrel{(2)}{{\cal T}}\!^{\alpha} = \frac{1}{3}{\cal T}\wedge\theta^{\alpha} \;,
\;\; {\cal T} = * (\theta_\alpha \wedge *{\cal T}^{\alpha} ) \;,\nonumber\\
\stackrel{(3)}{{\cal T}}\!^{\alpha} = \frac{1}{3}*({\cal P}\wedge \theta^{\alpha})\;,
\;\;{\cal P} = * (\theta_\alpha\wedge {\cal T}^{\alpha}) \;. \nonumber
\end{eqnarray}
Here the torsion trace 1-form ${\cal T}$ and the torsion pseudotrace 1-form
${\cal P}$ are introduced.
\par
It is convenient to use the auxiliary fields of 3-forms
$\eta_\alpha$, 2-forms $\eta_{\alpha\beta}$, 1-forms
$\eta_{\alpha\beta\gamma}$ and 0-forms
$\eta_{\alpha\beta\gamma\lambda}$,
\begin{eqnarray}
&&\eta_{\alpha} =  *\theta_{\alpha}\,,\quad \eta_{\alpha\beta} =
*(\theta_{\alpha}\wedge \theta_{\beta})\,, \nonumber\\
&&\eta_{\alpha\beta\gamma} = *(\theta_{\alpha}\wedge\theta_{\beta}
\wedge\theta_{\gamma})\,,\nonumber\\
&&\eta_{\alpha\beta\gamma\lambda} = *(\theta_{\alpha}\wedge
\theta_{\beta}\wedge \theta_{\gamma}\wedge \theta_{\lambda}) \,,
\end{eqnarray}
where $*$ is the Hodge operator.
\par
    In Weyl--Cartan space the following decom\-posi\-tion of the connection 1-form
is valid:
\begin{eqnarray}
&&\Gamma^{\alpha}\!_{\beta} = \stackrel{C}{\Gamma}\!^{\alpha}\!_{\beta} +
\Delta^{\alpha}\!_{\beta} , \nonumber\\
&&\Delta_{\alpha\beta} = \frac{1}{8}
(2\theta_{[\alpha}Q_{\beta ]} + g_{\alpha\beta} {\cal Q})\; ,\nonumber
\end{eqnarray}
where $\stackrel{C}{\Gamma}\!^{\alpha}\!_{\beta}$ denotes a connection
1-form of a Ri\-e\-mann--Cartan space $U_{4}$ with curvature, torsion and
metric compatible with the connection. This decomposition of the connection
induces the corres\-ponding decomposition of the curvature 2-form \cite{Mod},
\begin{eqnarray}
&&{\cal R}^{\alpha}\!_{\beta} = \stackrel{C}{{\cal R}}\!^{\alpha}\!_{\beta}
+ \stackrel{C}{{\rm D}}\Delta^{\alpha}\!_{\beta} +
\Delta^{\alpha}\!_{\gamma}\wedge \Delta^{\gamma}\!_{\beta} \nonumber \\
&& = \stackrel{C}
{{\cal R}}\!^{\alpha}\!_{\beta} + \frac{1}{4} \delta^{\alpha}_{\beta}{\cal R}
^{\gamma}\!_{\gamma} + {\cal P}^{\alpha}\!_{\beta}\;,  \nonumber\\
&&{\cal P}_{\alpha\beta} = \frac{1}{4}\Bigl ({\cal T}_{[\alpha}Q_{\beta ]}
- \theta_{[\alpha}\wedge \stackrel{C}{{\rm D}}Q_{\beta ]}\nonumber\\
&&+ \frac{1}{8}\theta_{[\alpha}Q_{\beta ]} \wedge {\cal Q}
- \frac{1}{16}\theta_{\alpha} \wedge \theta_{\beta}\, Q_{\gamma}Q^{\gamma}\Bigr )\; ,
\nonumber
\end{eqnarray}
where $\stackrel{C}{{\rm D}}$ is the exterior covariant differential with
respect to the Riemann--\-Car\-tan con\-nec\-ti\-on 1-form
$\stackrel{C}{\Gamma }\!^{\alpha}\!_{\beta}$,
$\stackrel{C}{{\cal R}}\!^{\alpha}\!_{\beta}$ is the Riemann--Cartan
curvature 2-form and
${\cal R}^{\gamma}\!_{\gamma} = {\rm d} \Gamma^{\gamma}\!_{\gamma} =
(1/2){\rm d}{\cal Q}$ is the Weyl segmental curvature 2-form.

\section{Perfect dilaton-spin  fluid as a source
         of Weyl--Cartan spacetime}

There are two dominated components of cosmolo\-gical medium in the present
epoch: dark energy and dark matter. According to our hypothesis \cite{CQG-03},
dark matter is realized as a perfect fluid with intrinsic degrees
of freedom, namely, the perfect dilaton-spin fluid.  Every particle of this fluid
is endowed with spin $S_{\alpha\beta}$ and dilaton charge $J$. These material
sources generate a Weyl--Cartan geometrical structure of spacetime.
\par
    The following Noether currents appear in the variational formalism \cite{Mod}:
the canonical energy-momentum 3-form $\Sigma_{\sigma}$,
\begin{equation}
\Sigma_{\sigma} = \frac{\delta {\cal L}_{fluid}}{\delta
\theta^{\sigma}} = p\eta_{\sigma} + (\varepsilon + p) u_{\sigma} u
+ n \dot{S}_{\sigma}\!_{\rho} u^{\rho} u, \label{eq:sigma}
\end{equation}
and the dilaton-spin momentum 3-form ${\cal J}^{\alpha}\!_{\beta}$,
\begin{equation}
{\cal J}^{\alpha}\!_{\beta} = - \frac{\delta{\cal
L}_{fluid}}{\delta \Gamma^{\beta}\!_{\alpha}} = \frac{1}{2}n \left
(S^{\alpha}\!_{\beta} + \frac{1}{4} J\delta^{\alpha}_{\beta}\right
)u.\label{eq:J}
\end{equation}
Here $\varepsilon$  is the internal energy density of the fluid,
${S}_{\sigma}\!_{\rho}$ is the specific (per particle) spin
tensor, $n$ is the fluid particles concentration, $p$ is the
hydrody\-namic fluid pressure, $u=u^{\alpha}\eta_{\alpha}$ is
the flow 3-form which corresponds the 1-form of velocity
$*u=u_{\alpha}\theta^{\alpha}$.
The `dot' notation for the tensor object $\Phi$ is introduced,
$\dot {\Phi}^{\alpha}\!_{\beta} = *(u\wedge {\rm D}
\Phi^{\alpha}\!_ {\beta})$.
\par
    In a Weyl--Cartan space the matter Lagrangian obeys the diffeomorphism
invariance which leads to the equation of motion of the perfect dilaton-spin
fluid in the form of a generalized hydrody\-na\-mic Euler-type equation.
The component of this equation along the 4-velocity yields the
energy conservation law along a streamline of the fluid \cite{Mod}:
\begin{equation}
{\rm d}\varepsilon = \frac{\varepsilon + p}{n}\, {\rm d} n\;. \label{eq:cons}
\end{equation}
\par
    The total Lagrangian density 4-form of the theory ${\cal L}$ is represented
as the sum of the Lagran\-gian density 4-forms of the gravitational
field ${\cal L}_{grav}$ and of the dilaton-spin fluid ${\cal
L}_{fluid}$,
\begin{equation}
{\cal L} = {\cal L}_{grav} + {\cal L}_{fluid}\;,
\end{equation}
where the Lagrangian density 4-form of the gravita\-tio\-nal field is
\begin{eqnarray}
&&{\cal L}_{grav} =2f_{0}\biggl (\frac{1}{2}{\cal R}^{\alpha}\!_{\beta}
\wedge \eta_\alpha\!^\beta + \frac{1}{4}\lambda
\,{\cal R}^\alpha\!_\alpha \wedge *{\cal R}^\beta\!_\beta \nonumber \\
&& + \varrho_1 \,{\cal T}^{\alpha} \wedge *{\cal T}_{\alpha}
+ \varrho_2 \,({\cal T}^{\alpha} \wedge {\theta}_{\beta}) \wedge
* ({\cal T}^{\beta}\wedge{\theta}_{\alpha}) \nonumber \\
&&+ \varrho_3 \,({\cal T}^{\alpha}\wedge {\theta}_{\alpha})\wedge * ({\cal T}
^{\beta}\wedge {\theta}_{\beta}) \nonumber \\
&&+ \xi \,{\cal Q}\wedge * {\cal Q} + \zeta \,{\cal Q}\wedge \theta^\alpha
\wedge *{\cal T}_\alpha - \Lambda \eta \biggr ) \nonumber\\
&&+ \Lambda^{\alpha\beta}\wedge \left ({\cal Q}
_{\alpha\beta} - \frac{1}{4}g_{\alpha\beta}{\cal Q}\right )\;. \label{eq:tlag}
\end{eqnarray}
Here $f_0 = 1/(2\mbox{\ae})$ ($\mbox{\ae} = 8\pi G $), $\Lambda$ is the
cosmolo\-gi\-cal constant, $\lambda$, $\varrho_1$, $\varrho_2$, $\varrho_3$,
$\xi$, $\zeta$ are the coupling cons\-tants, and $\Lambda^{\alpha\beta}$ is
the Lagrange multiplier 3-form with the properties,
$\Lambda^{\alpha\beta} = \Lambda^{\beta\alpha}$, $\Lambda^\alpha\!_\alpha = 0$.
\par
In (\ref{eq:tlag}) the first term is the linear Hilbert--Einstein Lagrangian
generalized to Weyl--Cartan space, the second term is the Weyl quadratic
Lagrangian. The Weyl 1-form ${\cal Q}$, in contrast to the Weyl's
classical theory, represents the gauge field, which is not related to
an electromagnetic field, which is pointed out in \cite{Ut}, \cite{Tuck}.
We call the Weyl 1-form ${\cal Q}$ a dilatation field (Weyl field). The term
with the coupling constant $\zeta$ represents the contact in\-ter\-ac\-ti\-on
of the dilatation field with torsion, which can occur in Weyl--Cartan space.
\par
The gravitational field equations in Weyl--Cartan spacetime can be obtained by
the varia\-ti\-nal procedure of the first order by varying the
Lagrangian (\ref{eq:tlag}) with respect to the connection 1-form
$\Gamma^{\alpha}\!_{\beta}$ ($\Gamma$-equation) and to the basis 1-form
$\theta^\alpha$ ($\theta$-equation) independently, the constraints on
the connection 1-form in Weyl--Cartan space being satisfied with the aid of
the Lagrange multiplier 3-form $\Lambda^{\alpha \beta}$ \cite{CQG-03}.
\par
Inclusion of a term with the Lagrange multiplier $\Lambda ^{\alpha \beta }$
into the Lagrangian density 4-form means that the theory is
considered in Weyl--Cartan spacetime from the very beginning
\cite{GC-99}--\cite{GC-03}. Ano\-ther variational approach has been developed in
\cite{Ob-Hel} where the field equations in Weyl--Cartan spacetime
have been obtained as a limiting case of the field equations of the
metric-affine gauge theory of gravity. These two approaches are not
identical in general and coincide only when $\Lambda^{\alpha\beta}$
is equal to zero as a consequence of the field equations.
\par
    Our variational method also differs from that of \cite{Tuck}, where a
variational procedure is constructed for obtaining field equations in empty spacetime,
and then material spinless sources are added to these field equations.
In contrast to \cite{Tuck}, we obtain the following consequences of the
$\Gamma$-equation,
\begin{eqnarray}
&&{\cal T} = \frac{3(\frac{1}{4} + \zeta)}{2(1 - \varrho_1 + 2\varrho_2 )}
{\cal Q}\; ,  \label{eq:con} \\
&&(1 -4\varrho_1 -4\varrho_2 -12\varrho_3 ) {\cal P}\nonumber \\
&&= - \frac{1}{2}
\mbox{\ae} n S^{\alpha\beta} u^\gamma \eta_{\alpha\beta\gamma}\;, \label{eq:P}\\
&&(1 +2\varrho_1 + 2\varrho_2 )\stackrel{(1)}{{\cal T}}\!_{\alpha}\nonumber \\
&&= - \frac{2}{3}
\mbox{\ae} n S_{\beta(\alpha}u_{\gamma)}\theta^\beta \wedge \theta^\gamma\; . \label{eq:T1}
\end{eqnarray}
\par
    Contracting the $\Gamma$-equation and using (\ref{eq:con}),
one finds an equation of Proca type for the Weyl 1-form \cite{CQG-03}
(compare with \cite{Tuck}, \cite{Ob-Hel}):
\begin{eqnarray}
*{\rm d} * \!{\rm d} {\cal Q} + m^{2} {\cal Q} = \frac{{\mbox \ae}}
{2\lambda} n J *\!u \,, \label{eq:dQ}\\
m^{2} = 16\frac{\xi }{\lambda } + \frac{3(\varrho _1 - 2\varrho_2 + 8\zeta
(1 + 2\zeta ))}{4\lambda (1 -\varrho_1 +2\varrho_2 )}\,.
\end{eqnarray}
Eq (\ref{eq:dQ}) shows that the Weyl field ${\cal Q}$, in contrast to Maxwell field,
possesses a non-zero rest mass and exhibits a short-range nature \cite{Ut,Arg,Tuck,Sola}.
\par
    Eqs. (\ref{eq:con}), (\ref{eq:P}) and (\ref{eq:T1}) solve the
problem of evaluation of the torsion 2-form. Using the algebraic field equations
(\ref{eq:P}) and (\ref{eq:T1}), the traceless and pseudotrace pieces of the torsion 2-form
are deter\-min\-ed via the spin tensor and the flow 3-form $u$ of the perfect dilaton-spin
fluid in general case. Using Eq (\ref{eq:con}), one can determine the torsion
trace 2-form via the dilatation field ${\cal Q}$, for which the differential
field equation (\ref{eq:dQ}) is valid, and the torsion trace 2-form
can propagate in the theory under consideration.

\section{\bf Modified  Friedmann--Lema\^{\i}tre equation}

Let us consider  a cosmological model with the Friedmann--Robertson--Walker
(FRW) metric with the scale factor $a(t)$,
\begin{equation}
{\rm d} s^{2} = \frac{a^{2} (t)}{1 - kr^{2}}{\rm d} r^{2} + a^{2} (t)r^{2}
{\rm d}\Omega^2 - {\rm d} t^{2}\;, \label{eq:RW}
\end{equation}
In this model the homogeneous and isotropic Universe is filled with a
perfect dilaton-spin fluid \cite{CQG-03}, which realizes
a model of dark matter with $J \not = 0$ in contrast to baryonic and
quark matter with $J = 0$. In \cite{CQG-03} it is shown that in a theory
with the Lagrangian (\ref{eq:tlag})  only spinless matter with $S_{\alpha\beta}=0$
can be a source of the gravitational field in a homogeneous and isotropic Universe with
the FRW metric (\ref{eq:RW}). In this case, the torsion 2-form consists
of only the trace piece. For the FRW metric (\ref{eq:RW}) also ${\rm d} {\cal Q} = 0$
identically, and one can derive the Weyl 1-form ${\cal Q}$ algebraically from
Eq (\ref{eq:dQ}).
\par
    For the FRW metric (\ref{eq:RW}) the continuity equa\-ti\-on ${\rm d} (nu)=0$
(${\rm d}$ is the exterior differentiation operator)
yields the matter conservation law $na^3 = N = {\rm const}$. As an equation
of state of the dilaton fluid, we choose the equation of state
$p = \gamma \varepsilon$, $0 \le \gamma \le 1$. Then integration of the
energy conservation law (\ref{eq:cons}) for the FRW metric (\ref{eq:RW}) yields
\begin{equation}
\varepsilon\, a^{3(1 +\gamma )} = {\cal E}_\gamma = {\rm const} \;, \qquad
{\cal E}_\gamma >0 \;. \label{eq:E}
\end{equation}
\par
    Variation of (\ref{eq:tlag}) with respect to the basis 1-form
$\theta^{\alpha}$ gives one more field equation ($\theta$-equation) with a
source in the form of the fluid canonical energy-momentum 3-form (\ref{eq:sigma}).
In \cite{CQG-03} we have decomposed the field $\theta$-equation into Riemannian and
non-Rie\-man\-ni\-an pieces. As a result, we can represent the field
$\theta$-equation as an Einstein-like equation
\begin{equation}
\stackrel{R}{R}_{\alpha\beta} - \frac{1}{2}g_{\alpha\beta}\stackrel{R}{R} =
 \mbox{\ae} \Bigl ((\varepsilon_{\rm {e}} +p_{\rm {e}} ) u_\alpha u_\beta +
p_{\rm{e}}g_{\alpha\beta}\Bigr ),  \label{eq:ein1}
\end{equation}
where $\stackrel{R}{R}_{\alpha\beta}$, $\stackrel{R}{R}$ are the Ricci tensor
and the curvature scalar of Riemann space, respectively,
$\varepsilon_{\rm e}$ and $p_{\rm e}$ are the energy density and pressure
of an effective perfect fluid:
\begin{eqnarray}
&&\varepsilon_{\rm{e}} = \varepsilon + \varepsilon_{\rm{v}} -
{\cal E}\left(\frac{n}{N}\right )^{2}\,,\nonumber\\
&&p_{\rm{e}} = p + p_{\rm{v}} - {\cal E}\left(\frac{n}{N}\right )^{2}\,, \nonumber\\
&& {\cal E} = \alpha\mbox{\ae}\left (\frac{JN}{2\lambda m^2}\right )^2\;,
\nonumber\\
&& \alpha =\frac{3\left (\frac{1}{4} +\zeta \right )^2}
{4(1 - \varrho_1 + 2\varrho_2 )} + \xi - \frac{3}{64}\;, \nonumber
\end{eqnarray}
and, moreover, $\varepsilon_{\rm v} = \Lambda /\mbox{\ae}$ and $p_{\rm v} =
-\Lambda /\mbox{\ae}$ are the energy density and pressure of vacuum with
the equation of state $\varepsilon_{\rm v} = -p_{\rm v} > 0 $.
\par
The field equation (\ref{eq:ein1}) yields the modified
Friedmann--Lema\^{\i}tre (FL) equation for the perfect dilaton fluid with
the equation of state $p = \gamma \varepsilon$,
\begin{eqnarray}
&&\left (\frac{\dot a}{a} \right )^2  + \frac{k}{a^2}\nonumber\\
&& = \frac{\mbox{\ae}}
{3a^6} \left ( \varepsilon_{\rm{v}} a^6 + {\cal E}_\gamma a^{3(1-\gamma)}
- {\cal E}\right )\,. \label{eq:fridd}
\end{eqnarray}
\par
    We put $k = 0$ in (\ref{eq:fridd}) in accordance with the modern
observational evidence \cite{nabl,Os-Perl,St}, which shows that the
Universe is spatially flat in the cos\-mo\-lo\-gi\-cal scale.
\par
    Another component of Eq. (\ref{eq:ein1}) has the form
\begin{equation}
\frac{\ddot a}{a}=\frac {\mbox{\ae} }{3a^6} \left [\varepsilon_{\rm{v}}a^6 -
\frac{1}{2}(1+3\gamma ){\cal E}_\gamma a^{3(1-\gamma)} + 2{\cal E} \right ]\!.
\label{eq:add}
\end{equation}

\section{Evolution scenario of the Universe with dilaton dark matter}

Our hypothesis consists in assuming that  the Universe evolution begins from
the superstiff stage, when the equation of stage of the dilaton fluid
is $\gamma = 1$. In this case the equation (\ref{eq:E}) yields
$\varepsilon\, a^6 = {\mathcal E}_1 = {\rm const}$. The FL equation
(\ref{eq:fridd}) reads
\begin{eqnarray}
\left (\frac{\dot a}{a} \right )^2  &=& \frac{\mbox{\ae}}{3a^6}
\left (\varepsilon_{\rm{v}}a^6 + {\mathcal E}_1 - {\cal E} \right )\nonumber\\
&=& \frac{\mbox{\ae}\varepsilon_{\rm{v}}}{3a^6} (a^6 -  a^6_{\rm min } )\;, \nonumber
\end{eqnarray}
and can be exactly integrated. The solution cor\-res\-pon\-d\-ing to the
initial data $t=0$, $a = a_{{\rm min}}$ reads \cite{CQG-03},
\begin{eqnarray}
&& a = a_{{\rm min}} (\cosh{\sqrt{3\Lambda}\,t})^{1/3}\;, \nonumber\\
&& a_{{\rm min}} = \left (\frac{\alpha {\mbox{\ae}}^2}{\Lambda}
\left (\frac{JN} {2\lambda m^2}\right )^2 - \frac{\mbox{\ae} {\cal E}_1}
{\Lambda}\right )^{1/6}\;. \nonumber
\end{eqnarray}
This solution describes the inflation-like stage of the evolution of the
Universe, which continues until the equation of state of the dilaton matter
changes and becomes no more superstiff.
\par
When a smooth jump of equation of state from $\gamma = 1$ to $\gamma = 1/3$
happens, the FL equation (\ref{eq:fridd}) describes a graceful  exit from the superrigid
stage to the radiation stage of the Universe evolution.
\par
Consider the radiation-dominated stage with $\gamma = 1/3$. In this case the Eq (\ref{eq:E})
yields $\varepsilon\, a^4 = {\mathcal E}_{1/3} = {\rm const}$. The modified Friedmann--Lema\^{\i}tre
equation (\ref{eq:fridd}) for the expanding Universe ($\dot a >0$) takes the form
\begin{equation}
\dot{a}=\sqrt{\frac{\ae\varepsilon_{\rm v}}{3}}\cdot\frac{1}{a^{2}} \left(a^{6}+\frac{\mathcal{E}_{1/3}}
{\varepsilon_{\rm v}}a^{2} - \frac{\mathcal{E}}{\varepsilon_{\rm v}}\right )^{1/2}\;. \label{eq:7}
\end{equation}
\par
    When $\dot{a} = 0$, an extremum of the scale factor is realized. In case $a\ll 1$, when
$\mathcal{E}$ is positive ($\alpha >0$) and sufficiently small compared the value of
$\mathcal{E}_{1/3}$, this extremum is approximately equal to
\[
a_{m1} \approx \left( \frac{\mathcal{E}}{\mathcal{E}_{1/3}}\right)^{1/2}=a_{1}\ll 1\;.
\]
This value realizes a minimum of the scale factor because from Eq (\ref{eq:add})
one can see that for this value the $\ddot{a}>0$.
\par
    In the limit $t\rightarrow 0$, $a\rightarrow a_{1}$ Eq (\ref{eq:7})
can be integrated giving
\[
a^{2}+a^{2}_{1}\ln\frac{a}{a_{1}}=a^{2}_{1} + 2\sqrt{\frac{\ae\mathcal{E}_{1/3}}{3}}\; t\;.
\]
This solution demonstrates a correct behavior $a \sim \sqrt{t}$ of the scale factor of
the Fried\-mann radiation-dominated stage at small (but not infinitesimal) time.
\par
When the radiation energy density becomes sufficient\-ly small compared to matter energy
den\-si\-ty, a matter-dominated stage begins with $\gamma = 2/3$. Then (\ref{eq:E})
yields $\varepsilon a^{5}= \mathcal{E}_{2/3}= {\rm const}$, and the modified FL equation (\ref{eq:fridd})
for $\dot a >0$ takes the form
\begin{equation}
\dot{a}=\sqrt{\frac{\ae\varepsilon_{\rm v}}{3}}\cdot\frac{1}{a^{2}} \left(a^{6}+\frac{\mathcal{E}_{2/3}}
{\varepsilon_{\rm v}}a - \frac{\mathcal{E}}{\varepsilon_{\rm v}}\right )^{1/2}\;. \label{eq:11}
\end{equation}
If $\mathcal{E}$ is sufficiently small compared to $\mathcal{E}_{2/3}$, a minimum of the scale factor
(when $\dot{a} = 0$ holds) in case $a\ll 1$ is approximately given by
\begin{equation}
a_{m2} \approx \frac{\mathcal{E}}{\mathcal{E}_{2/3}}=a_{2}\ll 1\;.
\label{eq:m2}
\end{equation}
\par
Eq (\ref{eq:add}) in case $\gamma = 2/3$ reads
\begin{equation}
\frac{\ddot a}{a}=\frac {\mbox{\ae} }{3a^6} \left (\varepsilon_{\rm{v}}a^6 -
\frac{3}{2}a{\cal E}_{2/3} + 2{\cal E} \right ), \label{eq:ad23}
\end{equation}
Using this equation, one can easily verify  that the value (\ref{eq:m2}) of the scale factor
corresponds to the condition $\dot a >0$.
\par
    In the limiting case $t\rightarrow 0$, $a\rightarrow a_{2}$ Eq (\ref{eq:11})
can be integrated with the solution,
\[
a^{5/2} + \frac{5}{6}a_{2}a^{3/2} = \frac{11}{6}a^{5/2}_{2} + \frac{5}{2}
\sqrt{\frac{\ae\mathcal{E}_{2/3}}{3}}\; t\;.
\]
This solution demonstrates a correct behavior $a \sim t^{2/5}$ of the scale factor of
the Friedmann matter-dominated stage at small (but not infinitesimal) time.
\par
    In the limit case $t\rightarrow \infty$, $a\rightarrow \infty$, Eq (\ref{eq:11})
has a de-Sitter-like solution with $\ddot a >0$,
\[
a = C\exp{\left (\frac{\Lambda}{3}\,t\right )}\;,\qquad C>0\:,
\]
where $C$ is an arbitrary positive constant. Thus an accelerating stage of the Universe
evolution is predicted.
\par
Equating the right-hand side of Eq (\ref{eq:ad23}) to zero, one can see that there are two points
of inflection of the scale factor plot \cite{CQG-03}. The first one has a very small value,
but the second one,
\[
a_{infl} \approx \left ( \frac{3\mbox{\ae}\mathcal{E}_{2/3}}{2\Lambda}\right )^{1/5}\;,
\]
corresponds to the modern era. This is the point, when the Friedmann expansion with deceleration
is replaced by an accelerated expansion, which agrees with the modern observational data
\cite{Perl}, \cite{Os-Perl}.
\par
    The last stage of expansion is a dust stage with $\gamma = 0$. In this case Eq (\ref{eq:E})
yields $\varepsilon a^{3}= \mathcal{E}_{0} = {\rm const}$, where $\mathcal{E}_{0}$
is the total mass-energy of dilaton matter of the Universe.
\par
The modified Friedmann--Lema\^{\i}tre equation (\ref{eq:fridd}) for the
expanding Universe ($\dot a >0$) takes the form
\begin{equation}
\dot{a} = \sqrt{\frac{\ae\varepsilon_{\rm v}}{3}}\cdot\frac{1}{a^{2}}\left ( a^{6}+\frac{\mathcal{E}_{0}}
{\varepsilon_{\rm v}}a^{3} - \frac{\mathcal{E}}{\varepsilon_{\rm v}}\right )^{1/2}\;. \label{eq:1}
\end{equation}
\par
When $\dot{a} = 0$, then we have a minimum of the scale factor,
\[
a^{3}_{m3} = -\frac{\mathcal{E}_{0}}{2\varepsilon_{\rm v}} +
\frac{1}{2}\sqrt{\frac{\mathcal{E}_{0}^{2}} {\varepsilon_{\rm v}^{2}} +
\frac{4\mathcal{E }}{\varepsilon_{\rm v}}}\;.
\]
If $\mathcal{E}_0$ is very large, one has
\[
a_{m3} \approx \left( \frac{\mathcal{E}}{\mathcal{E}_{0}}\right)^{1/3}=a_{0}\ll 1\;.
\]
\par
In this case Eq (\ref{eq:1}) can be exactly integrated with the solution
\begin{equation}
a^{3} + \frac{\mathcal{E}_{0}}{2 \varepsilon_{\rm v}} + \sqrt{a^{6}+\frac{\mathcal{E}_{0}}
{\varepsilon_{\rm v}}a^{3}  - \frac{\mathcal{E}}{\varepsilon_{\rm v}}} =
\mathcal{C} e^{\sqrt{3\Lambda}\,t}\;, \label{eq:3}
\end{equation}
where $C$ is an arbitrary constant. For the initial conditions $t=0$, $a=a_{m3}$,
the value of this constant is
\[
\mathcal{C} = \frac{1}{2}\sqrt{\frac{\mathcal{E}_{0}^{2}}
{\varepsilon_{\rm v}^{2}} + \frac{4\mathcal{E}}{\varepsilon_{\rm v}}}\;,
\]
and the solution (\ref{eq:3}) takes the form
\[
a \!=\! \left(\frac{\ae\mathcal{E}_{0}}{2\Lambda}\right)^{\frac{1}{3}}\!
\left( \sqrt{1+ \frac{4\Lambda\mathcal{E}}{\ae \mathcal{E}_{0}^{2}}}
\coth(\sqrt{3\Lambda}\;t) -1\right)^{\frac{1}{3}}.
\]
Another form of this solution is
\begin{eqnarray}
a = \biggl (\frac{\ae\mathcal{E}_{0}}{2\Lambda}\left(\coth(\sqrt{3\Lambda}\;t) -1\right) \nonumber\\
+ a^{3}_{m3} \coth(\sqrt{3\Lambda}\;t)\biggr )^{1/3}\;. \label{eq:6}
\end{eqnarray}
If one puts in (\ref{eq:6}) $a_{m3}=0$, there appears a cosmological monotonic model of $M_{1}$ type
\cite{Tol} which after an inflection point asymptotically turns into
the empty de Sitter universe as $t\rightarrow \infty$.

\section{Conclusions}

   Various non-standard cosmological theories lead to various modifications of the
Friedmann--Lema\^{\i}tre equation.
\par
In metric-affine gravity (MAG) one obtains a modified FL equation similar to (\ref{eq:fridd}),
but without a cosmological term \cite{Ob-Hel}. After analyzing this equation, the authors of \cite{Ob-Hel}
conclude that ``purely dilational matter amplifies gravitational attraction. In particular, it
accelerates rather then retards the possible collapse of a system.'' In this cosmological theory, the
analogue of our constant $\alpha$ is negative that ``corresponds to an additional effective
{\it at\-trac\-t\-ive} force dominating during the very early stages of evolution'' of the Universe. Recently
in \cite{Puet} the SN Ia supernovae data were analyzed  within this non-standard cosmological model with the
cosmological term added.
\par
In \cite{Tuck}, the similar modified FL equation (also without a cosmological term) was obtained in the
framework of the Einstein--Proca--matter system appearing from the Weyl--Cartan geometrical ap\-pro\-ach to
the gravitational theory. For the pressure-free dust case $\gamma=0$ it was shown by numerical methods
that this equation has both singular and nonsingular solutions.
\par
In \cite{Bar}, in the framework of version of D-brain cosmology on the boundary of anti-de Sitter space,
the modified FL equation similar to (\ref{eq:fridd}) appears with $\mathcal{E} \sim Q_{4+1}^2$, where $Q_{4+1}$
corresponds to an ``electric charge'' in (4+1)-dimen\-sional sense.
\par
    In our theory, the set of equations (\ref{eq:fridd})--(\ref{eq:add}) describes a nonsingular model
of evolution of the Universe starting from an inflation-like stage (for superstiff equation
of state), passing through radiation-dominated and matter-dominated dece\-le\-ra\-t\-ing stages and turning into
a post-Friedmann accelerating era.

\renewcommand{\refname}{References}

\end{document}